\documentclass[12pt,manuscript]{aastex}

\begin{document}

\title{The Carnegie Astrometric Planet Search Program}

\author{Alan P.~Boss, Alycia J.~Weinberger, and Guillem Anglada-Escud\'e}
\affil{Department of Terrestrial Magnetism, Carnegie Institution of
Washington, 5241 Broad Branch Road, NW, Washington, DC 20015-1305}

\author{Ian B.~Thompson, Gregory Burley, and Christoph Birk}
\affil{Carnegie Observatories, 813 Santa Barbara Street,
Pasadena, CA  91101-1292}

\author{Steven H.~Pravdo and Stuart B.~Shaklan}
\affil{Jet Propulsion Laboratory, California Institute of Technology,
4800 Oak Grove Drive, Pasadena, CA  91109}

\author{George D.~Gatewood}
\affil{Allegheny Observatory, University of Pittsburgh, 159 Riverview Ave.,
Pittsburgh, PA 15214}

\author{Steven R.~Majewski and Richard J.~Patterson}
\affil{Dept. of Astronomy, University of Virginia, 530 McCormick Road, 
Charlottesvile, VA 22903-0818}

\begin{abstract}
 
 We are undertaking an astrometric search for gas giant planets and
brown dwarfs orbiting nearby low mass dwarf stars with the 2.5-m 
du Pont telescope at the Las Campanas Observatory in Chile. We have 
built two specialized astrometric cameras, the Carnegie Astrometric 
Planet Search Cameras (CAPSCam-S and CAPSCam-N), using two
Teledyne Hawaii-2RG HyViSI arrays, with the cameras' design having 
been optimized for high accuracy astrometry of M dwarf stars. We 
describe two independent CAPSCam data reduction approaches and present a 
detailed analysis of the observations to date of one of our target stars, 
NLTT 48256. Observations of NLTT 48256 taken since July 2007 with 
CAPSCam-S imply that astrometric accuracies of around 0.3 milliarcsec per 
hour are achievable, sufficient to detect a Jupiter-mass companion 
orbiting 1 AU from a late M dwarf 10 pc away with a signal-to-noise 
ratio of about 4. We plan to follow about 100 nearby (primarily within 
about 10 pc) low mass stars, principally late M, L, and T dwarfs, for 10
years or more, in order to detect very low mass companions with orbital
periods long enough to permit the existence of habitable, Earth-like
planets on shorter-period orbits. These stars are generally too faint 
and red to be included in ground-based Doppler planet surveys, which
are often optimized for FGK dwarfs. The smaller masses of late M dwarfs
also yield correspondingly larger astrometric signals for a given 
mass planet. Our search will help to determine whether gas giant 
planets form primarily by core accretion or by disk instability
around late M dwarf stars.

\end{abstract}


\section{Introduction}

 There are only 21 known G stars within 10 pc of the sun, but at least 
239 M dwarfs (Henry et al. 2006), stars with masses in the range
of 0.08 to 0.5 $M_\odot$. Given this extreme imbalance in the 
numbers of the closest stars, M dwarfs are a natural choice for 
astrometric planet searches, where closeness is the primary virtue.
Young M dwarf stars appear to have protoplanetary disks similar to those 
around T Tauri stars of higher mass (e.g., Andrews \& Williams 2005) and 
perhaps even longer lived (Carpenter et al. 2006). Hence there is no obvious
reason to believe that low mass stars should not be able to form 
planetary systems in much the same manner as their somewhat more 
massive siblings. In fact, radial velocity and microlensing searches 
have begun to discover planetary companions (gas giants and hot or 
cold super-Earths, respectively) to M dwarf stars in some abundance 
(e.g., Marcy et al. 1998, 2001; Butler et al. 2004; Bond et al. 2004; 
Rivera et al. 2005; Bonfils et al. 2005; Udalski et al. 2005), and
a candidate gas giant planet orbiting an M dwarf (VB10) has been 
astrometrically discovered (Pravdo \& Shaklan 2009). In addition,
brown dwarfs should be more frequent companions to M dwarfs than to
G dwarfs, given the smaller mass ratio involved (e.g., 
Joergens 2008; Jao et al. 2009). These brown dwarf 
companions will be considerably more astrometrically detectable 
around M dwarfs than gas giant planet companions. We are focusing
our astrometric search on late M dwarfs and even fainter stars 
(L and T dwarfs), targets that are not generally included in 
radial velocity surveys.

 Our astrometric search will aid in the determination of which of two 
competing mechanisms for gas giant planet formation dominates by searching 
for giant planets around M dwarfs. Wetherill (1996) found that Earth-like 
planets were just as likely to form from the collisional accumulation of 
solids around M dwarfs with half the mass of the Sun as they were to form
around solar-mass stars. Boss (1995) studied the thermodynamics of
protoplanetary disks around stars with masses from 0.1 to 1.0 $M_\odot$,
and found that the location of the ice condensation point only
moved inward by a few AU at most when the stellar mass was decreased
to that of late M dwarfs. In the core accretion model of giant
planet formation, this implies that gas giant planets should be
able to form equally well around M dwarf stars, and perhaps at
somewhat smaller orbital distances. However, the longer orbital
periods at a given distance from a lower mass star mean that core
accretion may be too slow to produce Jupiter-mass planets 
around M dwarfs before the disk gas disappears (Laughlin et al. 2004;
Ida \& Lin 2005).
In the competing disk instability model for gas giant planet formation,
calculations for M dwarf protostars (Boss 2006) have shown that 
M dwarf disks are capable of forming gas giant protoplanets rapidly.
Gas giant planets formed by disk instability for host protostars
with masses of both 0.5 $M_\odot$ and 0.1 $M_\odot$ (Boss 2006), spanning
nearly the entire range of M dwarf masses, with no indication that
the process would not continue to operate for even lower mass dwarfs.
Hence a search for gas giant planets orbiting M, L, and T dwarfs should be a 
valuable means for determining if disk instability is able to form
giant planets in significant numbers around these dwarfs, as core accretion 
seems to be ruled out for such very low mass stars and brown dwarfs.

 In this paper, we describe the Carnegie Astrometric Planet Search 
Cameras (CAPSCam-S and CAPSCam-N), which are the centerpieces of our 
efforts. We also present data for one target field from the first two 
years of observations with CAPSCam-S, and use these observations to estimate 
the short- and long-term astrometric accuracy of CAPSCam-S on the du Pont 
telescope. 

\section{Carnegie Astrometric Planet Search Cameras}

 One of the main motivations for this program is to take advantage of the
Carnegie Institution's du Pont telescope for use in the search for 
extrasolar planets and brown dwarf stars. The Las Campanas Observatory 
(LCO) is located at an elevation of $\sim$ 7,200 feet (2200 meters) 
in the foothills of 
the Andes mountains near La Serena, Chile. The median free-air seeing at Las
Campanas is about 0.6 arcseconds (Persson et al. 1990), so the observatory
is quite well-suited for high precision, ground-based astrometry. The 
du Pont f/7.5 Cassegrain telescope, with its 2.5-m aperture and large 
isokinetic patch, is an excellent choice for our astrometric program. 
 
 Astrometric planet searches with CCDs typically suffer from a 
severe brightness contrast between the target and reference stars,
especially for G dwarf targets. Even a mid-M dwarf within 10 pc
has a V magnitude of $\sim$ 10 to 12, whereas good astrometric
reference stars lying at distances of hundreds of pc to a few
kpc have V magnitudes of $\sim$ 15 to 18 or more. CCD cameras
will thus saturate on the target star long before sufficient
photons from the references stars are collected.

 In 2004 we were awarded a grant from the NSF Advanced Technology and 
Instrumentation Program and Carnegie Institution matching funds to build a 
state-of-the-art astrometric camera to solve this bright target star problem. 
CAPSCam uses a Hawaii-2RG HyViSI hybrid array that allows the definition of an 
arbitrary guide window, which can be read out (and reset) rapidly, repeatedly, 
and independently of the rest of the array. This guide window is centered 
on our relatively bright target stars, with multiple short exposures to 
avoid saturation. The rest of the array then integrates for prolonged 
periods on the background reference grid of fainter stars. This can 
dramatically extend the dynamic range of the composite image. 
This HyViSI detector is the heart of the CAPSCam concept.

 The Hawaii-2RG array is a three-side buttable, silicon-based hybrid 
focal plane array, with Si PIN photodiodes indium-bump-bonded to a 
CMOS readout multiplexer. The active, light-sensitive area of the array 
is 2040 x 2040 active 18 um pixels, surrounded on each side (window 
frame style) by 4 rows and columns of reference pixels, 
for a total of 2048 x 2048 array elements (Bai et al. 2003). However,
in order to increase the dynamic range of CAPSCam-S, we have
set the bias voltages at a level such that an addtional 3 rows and
columns of pixels are affected by the bias in the reference pixels,
reducing the effective active imaging pixels to 2034 x 2034. The
gain in dynamic range more than makes up for this small loss in
imaging area.

 In our configuration, the detector operates with four output 
channels at 125 kHz. Each channel is a 2048 x 512 stripe of the array. 
The detector also has a guide window (GW) mode, with programmable 
size and location, which is read out by one of the four analog to digital 
converters. This mode allows a selected subarray to be read, 
without disturbing the main array (the full frame, or FF). A few simple 
clock signals and a serial interface are used to select and control 
the various detector array operations. The readout timing is 8 
microseconds per pixel. 

 We purchased two Teledyne (formerly Rockwell) HyViSI arrays, one science grade 
and one engineering grade. The science grade array is mounted in CAPSCam-S, 
and has a dark current of less than 0.1 electron/s/pixel and a read noise of 
12.5 electrons. The gain is 2.1 electrons per data number (DN), and the linear 
full well is 130,000 electrons. The pixel size is 0.194 arcsec, while the 
field of view of CAPSCam-S is 6.63 arcmin by 6.63 arcmin. 
The engineering grade array is comparable in performance to the science grade 
array, except with a read noise of 28 electrons. The engineering grade 
array has been used to build a second camera, CAPSCam-N, for use in testing 
in the laboratories in Pasadena and for use in a northern 
hemisphere planet search on the Mt. Wilson 2.5-m telescope (see Figure 1).

 The detector is mounted in our standard single-chip rectangular 
aluminum housing, 
coupled to an IR-Labs ND-2 cryostat. The control electronics box is mounted to 
the side of the dewar (Figure 2). The system is liquid nitrogen cooled, and has 
a hold time of approximately 13 hours.

 The CAPSCam dewar window serves as the system passband filter, with 
a wavelength range of about 800 to 930 nm. Figure 3 illustrates the 
throughput of CAPSCam as a result of the combination of the Hawaii-2RG 
array and the filter/window. The filter/window 
was made with a multilayer coating on a 90 mm diameter 
$\lambda$/30 (peak to valley) fused-silica window from Barr Associates.

 Two of us (GB and IBT) designed and built the electronics boards, and assembled
both CAPSCams in our laboratories in Pasadena. The detector controller is a 
compact, 4-channel digital signal-processor-based system. It is a modified 
version of our BASE ccd control electronics, which is described in detail on the
OCIW website. Tables \ref{table_readout} and \ref{table_exposure} present more details about the CAPSCam read-out scheme,
including the GW, FF, and overhead readout times.

 One would normally operate the array without a shutter, using the built-in
electronic reset function at the beginning of each FF or GW exposure. However,
this would result in different sampling of the atmospheric turbulence by the
wavefronts from the GW and FF stars, as the FF would continue to integrate 
during the time used for resetting and reading the GW. Hence
a mechanical Uniblitz shutter and driver was purchased from 
Vincent \& Associates, in order to synchronize the sky time 
for both the guide window and the full frame of the array. The 
optical path then consists solely of the primary and secondary mirrors,
the shutter, the Barr filter/window, and the Teledyne array, eliminating 
optical distortions caused by any additional components.

 CAPSCam-S camera was installed on the du Pont telescope in March 
2007 (Figure 2) and has been in operation ever since,
with several improvements having been made in the meantime. 
CAPSCam-S is controlled by a software graphical user interface 
(GUI) written by one of us (CB), which allows for convenient 
operation and monitoring of all of its functions. A detailed users 
manual for CAPSCam-S with a description of the GUI windows
can be found on the Carnegie Observatories web pages.

 The minimum integration time for the guide window is 0.2 sec, allowing  
CAPSCam to handle target M dwarf stars typically as bright as a V  
magnitude of $\sim$ 12 or an I magnitude of $\sim$ 9 without  
saturating the array.  Saturated pixels show persistence at the level  
of a few hundred DN, which can last for several hours or more, even  
with repeated flushing of the array by taking dark frames. As a  
result, we try to avoid bright stars anywhere in the image. However,  
the low level of the persistence, and its confinement to a few pixels  
at the cores of saturated stars, means that it does not interfere with  
accurate astrometry of stars elsewhere in the field.

\section{Target Star Sample Selection}

 Our top priority sample consists of 44 stars and brown dwarfs closer 
than 10 pc with spectral types later than M2.5, with another 26 dwarfs from 
10 pc to 20 pc, for a total of 70 targets with known parallaxes within 
20 pc. The majority of these targets have spectral type M5.5 or later.
For observability, we require that the targets not saturate in our 0.2 s  
minimum integration time, i.e., have I magnitudes greater than $\sim$ 8.5. 
Only three of our targets have V magnitudes less than 12, which is a 
typical limiting magnitude for radial velocity surveys, even on large 
telescopes.
 
 We restricted our sample to targets south of +16 deg declination in  
order to minimize airmass effects at the latitude of Las  
Campanas (-29 deg). Known close binary star systems have been  
avoided. We have supplemented our list with a number of M,  
L, and T dwarf targets from the 2MASS catalogs, which do not have  
parallax determinations yet. 10 have spectrophotometric parallaxes 
of 20 pc or less, 10 appear to lie beyond 20 pc, while 15 do not
have even a spectrophotometric parallax. The very faintest L and T dwarfs
(I $\sim$ 18 or more) have been dropped from consideration.
After determining their parallaxes, those stars with distances 
greater than about 20 pc will also be dropped from  
further consideration, unless they show evidence for possibly having  
detectable companions. We intend to follow the best $\sim$ 100 targets
for a decade or longer with CAPSCam-S on the du Pont.

\section{CAPSCam Data Analysis Pipeline}

 We present here a brief description of the astrometric data
pipeline developed at DTM specifically for CAPSCam. More complete
details about the data pipeline and algorithms employed will be given in 
a forthcoming publication (see Anglada-Escud\'e et al., in preparation). 
We have also analyzed CAPSCam data with a completely separate data analysis
pipeline developed by another of us (SBS), which has been used
extensively in the Stellar Planet Survey (STEPS) astrometric
program in the northern hemisphere on the Palomar 5-m telescope
(Pravdo \& Shaklan 1996). While the results obtained from the two approaches
contain significant differences, these differences can be understood in terms
of the different algorithms and data processing approaches used in the
two pipelines.

 CAPSCam data processing consists of two major steps: source
extraction (or one night processing) and the astrometric iterative
solution.

\subsection{One Night Processing}

 A chosen image is used to create an astrometric template of the field for
a given night. Some initial rejection of objects is done based on the roundness
(ratio of the full-width-half-maximum [FWHM] of the point spread function
[PSF] in the X and Y directions) and convergence properties of trying to
find an initial rough estimate for the photocenter of each object 
(centroid). The 20 brightest objects (typically stars) are identified
in every image and the rest of the objects are then located
using the astrometric template. A fine centroiding algorithm is
then applied and a catalog with the subpixel positions of all of the
objects is obtained for all of the images of the field for a given night. 

 The centroiding algorithm consists of binning the PSF in the X and Y
directions separately and fitting a one-dimensional PSF profile in each
direction. Several different window sizes (apertures) are used on 
each object and the centroid position is found by averaging over 
all the apertures tried. This scheme has been compared to 
two-dimensional (2D) approaches (e.g., 2D Gaussian PSF fitting), and
found to provide the most robust centroid determinations. The centroid
determinations are very stable numerically, give the smallest scatter, 
and are insensitive to the discreteness of the sampling of the PSF. 

 When all the images for a given night are
processed, the relative positions of all of the stars are compared,
and the resulting scatter is used to estimate the centroid uncertainties for
each star. Since we observe with telescope ditherings of $2^{''}$ 
in both X and Y, bad pixels and other Hawaii-2RG defects will move 
significantly and can be easily removed at this point. A final filtering 
of bad pixels and detector defects is then done to produce the final plate
catalogs. The processing of each image generates a \textit{plate catalog},
which contains a list of the X-Y-centroids and their associated
uncertainties. One astrometric epoch typically then consists of 
between 20 to 80 plate catalogs obtained in a given night.

\subsection{Astrometric Iterative Solution}

 Once images from a given field have been obtained on different nights
spread over a long time baseline, an astrometric solution can be
obtained and used to derive the positions, proper motions, and 
parallaxes of all the stars in each target field.

 The astrometric solution is an iterative process. An initial
catalog of positions is generated from a given plate, a
transformation is applied to each plate catalog to match the
initial catalog, and the apparent trajectory of each star is then fitted to
a basic astrometric model. The initial catalog is updated with new
positions, proper motions, and parallaxes and a subset of well-behaved
stars is selected to be used as the reference frame.
The selection of the reference stars is based on the number of
successful observations and the median of the root-mean-square (RMS) of the
residuals for the 100 brightest objects. A typical field contains around 40--50 of
such stars. This process is then iterated a small number of times using only the
reference stars for calibration purposes. The convergence of the astrometric solution
is monitored by following the average RMS of the reference frame stars.

 Since the number of targets followed and images generated by the CAPSCam
planet search effort is large, the astrometric solution process has 
been designed to be fully automatic. Albeit structurally simple, 
some steps in this processing are algorithmically complex, especially 
those related to cross-matching, reference star selection, calibration 
weighting, astrometric model selection, etc. A more complete description 
of each of these steps will be given in Anglada-Escud\'e et al. (in
preparation).

 The initial catalog is matched to the NOMAD catalog (Zacharias et 
al. 2005), which contains USNO-B1 and 2MASS positions and colors
of the brighter objects. This initial matching is required to
better constrain the field rotation and plate scale, and to obtain the
resulting astrometric positions in meaningful sky coordinates.

 The final product is then an astrometric catalog containing five 
fitted astrometric parameters (effectively X and Y, the proper motions
in X and Y, and the parallax), information about the number of 
observations employed, the RMS of the residuals per epoch, 
and the reduced $\chi^2$ of the solution for each star:

\begin{eqnarray}
\bar{\chi}^2 = \frac{1}{2 N_{epochs} - N_{pars}}\sum^{N_{epochs}}_i
\left[\frac{\left(x^i_{obs}-x_{model}\right)^2}{\sigma_i^2} +
\frac{\left(y^i_{obs}-y_{model}\right)^2}{\sigma_i^2}\right],
\end{eqnarray}

\noindent where $N_{epochs}$ is the number of epochs, $N_{pars}$ is 
the number of model parameters to be fit, $x^i_{obs}$ and 
$y^i_{obs}$ are the measured positions of the star in a local 
coordinate system, $x_{model}$ and $y_{model}$ are the ones predicted
by the best fit model, and the summation is over all epochs.
Each observation is weighted using its own standard deviation $\sigma_i$.

 Values of $\bar{\chi^2}$ greater
than unity typically indicate the presence of uncalibrated systematic
errors. These can occur for the target star as well as for the 
reference frame stars. The amount of this uncalibrated systematic error is
obtained for each iteration by adding an error in quadrature to
the estimated uncertainties derived from the calibration step
until an effective $\bar{\chi}^2 = 1$ is obtained. This
guarantees that the uncertainties in the parameters
from the astrometric least squares solution are more realistic
than the ones obtained using only the intra-night scatter.
Despite the fact that part of any systematic error may come from chromatic
effects, stars with less extreme colors than the red dwarf target stars 
show similar residuals (i.e., RMS deviations and $\bar{\chi^2}$) 
pointing to other sources of systematic errors related to the mechanical
and optical stability of the telescope and detector on the scale of
a few pixels. CAPSCam calibration observations are ongoing 
to try to clarify and identify the sources of such night-to-night and
long-term systematic errors, which should add more or less in quadrature, 
with the hope of eventually reducing or eliminating these errors 
whenever possible. Systematic error sources include dome seeing, 
dust on the optics, minute changes in the optical alignment 
(depending on the temperature, humidity, or gravity angle), and
the true effective photocenters of the individual pixels on the 
Hawaii-2RG.

 Geometric calibration in the CAPSCam data pipeline 
accounts for differential optical aberrations with
respect to the plate used as the initial catalog. We note that as long
as we deal with the same instrument, only the time dependent part of the
optical aberrations is relevant. This includes both atmospheric and
instrumental-related optical distortions. The CAPS pipeline permits
specifying the order of the polynomials to be used in the calibration. 
The zero order polynomial corresponds to a translational shift while
the first order polynomial corrects for a
small rotation and a shear. In the language of Zernike polynomials (see Noll
1975, his Table 1), the second order polynomials account for defocus 
and astigmatism. While the accuracy of the astrometric solution improves
significantly if second order polynomials are used, we find that there is no
significant improvement when using the third order ones as well; i.e., 
the third order Zernike aberration (coma) changes very little over the 
timespan of these CAPSCam-S observations.

\section{Differential Chromatic Refraction}

 Ideally, astrometric observations are taken as the target star
passes the meridian, in order to minimize atmospheric seeing
effects and differential chromatic refraction (DCR, e.g., Pravdo \&
Shaklan 1996). Uncalibrated DCR leads to systematic errors because
the photocenters of the target and reference stars will be
refracted differently as the air mass changes. Pravdo \& Shaklan
(1996) found that DCR could be calibrated to about 0.13 milliarcsec
for observations within 1 hour of the meridian and 45 degrees of
the zenith with the Palomar 5-m. We intend to minimize the effects of DCR
by using the same calibration technique for the du Pont and expect
to be able to remove DCR to a level similar to that found to be
possible at Palomar. The CAPSCam spectral bandpass has a FWHM of
about 100 nanometers, centered on 865 nanometers, which also limits DCR 
effects for our red target stars and typically red reference stars.

 In our first four years of du Pont observations (2003-2006), we used the Tek5 
CCD camera to take Washington+DDO51 photometry (Geisler 1986; Majewski et al. 
2000) of over 250 likely target stars (selected in part from Reyle \&
Robin 2004; Vrba et al. 2004; and Golimowski et al. 2004) and their
reference stars, which is needed in order to remove the effects of
DCR and to characterize the luminosity classes of the reference
stars (i.e., distant giants are preferred and are identifiable
with these filters; Majewski et al. 2000). Essentially all of the
prospective target stars have sufficient reference stars within
the 6.63 arcmin by 6.63 arcmin field of view of CAPSCam. We have
also added Johnson B,V photometry (Johnson \& Morgan 1953) for most fields, 
necessary for obtaining absolute parallaxes through reddening corrections. 
Hence, the characterization 
phase of our search is largely finished, though the occasional addition
of new target fields to the planet search will require further 
Tek5 runs to obtain their colors.

 At present, we apply a prototype of DCR correction in our CAPSCam data
analysis pipeline which can be turned on or off. It is based on the
colors from the NOMAD catalog (B,V from USNO-B1 and J,H,K from 2MASS).
Ultimately, the chromatic corrections will be based on the Tek5 photometric
determinations; this process is currently in development. Even
with our crude DCR corrections we already see a distinct improvement of the 
quality of our astrometric solutions. Observations of a few target 
fields followed for about 6 hours through airmasses ranging from 1 to 3 
are being used to estimate the DCR effects as a function of the TeK5 colors.

\section{First Results with CAPSCam-S}

 The key objective of the CAPSCam program is achieving and
maintaining an astrometric accuracy significantly better than a
milliarcsecond for a decade or longer.  The natural plate scale
for CAPSCam-S on the du Pont is 0.194 arcsec/pixel, a scale
that allows us to avoid introducing any extra optical elements into
the system that would produce astrometric errors. We take multiple
exposures with CAPSCam-S, typically 60 seconds for the full frame,
with small variations (2 arcsec) of the image position (dithering),
in order to average out uncertainties due to pixel response 
non-uniformity. We typically spend about one hour per field for
each epoch. The ultimate goal is to achieve a precision of about
0.25 milliarcsec, and perhaps as low as 0.15 milliarcsec, which
is the level of the atmospheric noise in one hour found by Pravdo
\& Shaklan's (1996) Palomar 5-m study. Here we summarize our results
to date.

 The goal of this first paper is limited to giving concrete evidence 
of the astrometric performance of CAPSCam-S. We do not discuss
important issues related to the choice of the reference frame or to
zero--point parallax and proper motion corrections. Even still, 
it is remarkable that the proper motions obtained for the objects 
in the studied target field agree fairly well with those given 
in the USNO-B1 catalog (Monet et al. 2003), especially the reference
stars.

\subsection{NLTT 48256}

 We present here the performance of CAPSCam-S on Field 453 (Figure 4),
which contains the target star NLTT 48256 (also known as LP 813-23 and
as 2MASS J19483753-1932140). The target star is an M dwarf about which 
little is known. The field is rich in background objects ($\sim 400$), 
a few of which are brighter than the target. The previously reported 
properties of this star are given in Table \ref{table_nltt48256}. From the photometry, it 
is clear that it is a very red star ($V-K \sim 6.3$), most likely with 
a spectral type of M5V-M6V. Six epochs spread out over a 2 year period 
are presented here for this field (July 2007--June 2009, see Figure 5), 
so the parallax and the proper motion can be
decoupled (for which only 3 epochs are required) and a preliminary 
discussion can be presented of the accuracy of the obtained solution,
at least in a statistical sense. Table \ref{table_obs} gives
the observing log information about the CAPSCam-S observations that
were included in the analysis that follows.

 NLTT 48256 is faint enough for CAPSCam-S to use FF imaging without
the need for the GW. The analysis that follows therefore is largely 
based on FF data for this field. However, we have also taken
GW data on Field 453 in June 2009, with the GW centered on the target 
star. Forty images were obtained on two separate nights. The resulting 
astrometric accuracy of the combined nights is shown in Figures 5 and 6,
though the GW epoch was not included in the solutions given in Tables 
\ref{table_solution} and \ref{table_shifts}. Figures 5 and 6 show 
that usage of the GW mode does not 
introduce a significant bias into the astrometric measurements, and 
provides a precision comparable to that of the FF mode.

\subsection{Achromatic solution}

 No color correction is applied in this first case. After excluding 
poorly-behaved stars in the first iteration, the reference frame still
contains 39 objects that appeared in at least $90\%$ of the frames.
These 39 objects define a robust reference frame with a median RMS per
epoch of 1.1 milliarcsec (mas). The astrometry of the poorly-behaved 
stars is also obtained, but they are not used in the calibration matching step.

 The RMS of the solution for the target star is $0.38$ mas/epoch, which
is a proxy for the long-term stability of the instrument. Since our
desired target star accuracy is $\sim 0.25$ mas/epoch, evidently further work 
will be required to try to reach this figure. Using the method described 
in the previous sections, we estimate the amount of systematic error, 
finding that it is $0.3$ mas/epoch. Clearly the systematic error is a 
significant source of the scatter in the residuals (see Figure 5, 
bottom panels). Similar values are achieved for a 
significant number of reference stars in the field, providing a 
robust double-check on the quality of the entire astrometric 
solution. The residuals appear to be somewhat larger in Dec.
than in R.A., and we are investigating possible sources of this
apparent difference.

 The target star does not show any excess in the post-fit residuals when
compared to other stars in the field with similar magnitude. In fact, the RMS
of the target star is slightly smaller than the average. Since the target star is 
located at the center of the field, where the optical distortions are less severe, 
this is not totally unexpected. 

 The final astrometric solution for the target star is given in
Table \ref{table_solution}. The proper motion and parallactic 
motion of NLTT 48256 are displayed in Figures 5 and 6. 
The resulting parallax is $\sim 17$ mas, which puts the target
star at a distance of $58$ pc, considerably beyond our distance
cut-off of about 10 pc. However, this star will be kept in the 
observing program in order to monitor the long-term stability 
of CAPSCam-S and to support our search for sources of systematic 
errors. Claims for astrometic planet detections are best supported
by observations showing that other target stars are not being
similarly perturbed (so-called ``flat-liners'', R. P. Butler, 
personal communication), so Field 453 will continue to provide an 
invaluable check on the short- and long-term astrometric accuracy 
of CAPSCam-S.

\subsection{Chromatic solution}

 The same procedure as above has been applied using $R-J$ color as a
variable in the calibration step, which is the color most closely related
to the slope of the spectral energy distribution in the CAPSCam working
band. The number of useful reference frame stars drops to 35 in
this case because only stars with known R and J colors are used. 
The RMS of the astrometric solution for the target star decreases 
to $0.35$ mas/epoch, showing a slight improvement in the accuracy. 
We note that the parallaxes determined with and without the chromatic
correction are incompatible at the several $\sigma$ level. This is
caused by the correlation of the DCR with the parallax factor,
which introduces a small bias into the parallax estimation. Since
our current version of the DCR correction is only a prototype, we
expect a small but significant increase in the accuracy and a
better decoupling of the true parallax from color dependent
effects once we are able to derive a solution with the full Tek5
photometric colors.

\subsection{Search for periodic signals}

 Once the main astrometric solution is finished, we can fit the plate 
motion of the target star with an astrometric model including a 
Keplerian component. However, with effectively only five epochs 
(i.e., 10 measurements of either X or Y), there is not enough 
information to solve for a fully Keplerian orbit 
plus the astrometric solution. Hence, this exercise should
be considered purely as an academic one. We can then run a Least
Squares periodogram routine, which consists of fitting for each 
orbital period $P$ sampled a linearized astrometric solution 
with a circular orbit in an arbitrary orientation, i.e.,

\begin{eqnarray}
X_{\alpha} &=& X_0 + \mu^*_\alpha \left(t-t_0\right) +
\Pi p_\alpha\left(t\right) + 
A \sin 2\pi/P + B \cos 2\pi/P \\
Y_{\delta} &=& Y_0 + \mu_\delta\left(t-t_0\right) + 
\Pi p_\delta\left(t\right) + 
C \sin 2\pi/P + D \cos 2\pi/P
\end{eqnarray}

\noindent where the offsets $X_0$ and $Y_0$, proper motions
$\mu^*_\alpha$ and $\mu_\delta$, parallax $\Pi$ and the orbital coefficients
A,B,C and D are solved simultaneously for each test period $P$. The 
functions $p_\alpha(t)$ and $p_\delta(t)$ are called the parallax factors. They
are the projections of the parallactic motion in R.A. and Declination in the 
direction of the star. The barycentric instant of observation 
is $t$ and the reference epoch at which the astrometric parameters are 
defined is $t_0$.

In order to maximize the sensitivity of our planet search, we have developed 
a Least Squares periodogram approach that may perform better than other
proposed methods, such as the Joint Lomb-Scargle periodogram proposed by
Catanzarite et al. (2006). Our Least Squares minimization solves 
simultaneously for the signal, parallax, proper motion, 
and two small offsets in each direction. All of these parameters are intrinstic 
to the astrometric measurements, and can correlate spuriously with the sampling 
cadence and the true signal, leading to incorrect identification of candidate 
periods. The Least Squares approach remains linear in all the free parameters 
if only astrometric data is involved (see, e.g., Pourbaix 1998), which makes it computationally efficient and numerically well-behaved. Black \& Scargle (1982) were the first
to point out that the coupling between the reflex signal and proper motion
will lead to underestimates of both the period and the amplitude of the signal,
even when the data span the entire period of the signal. We go one step further,
and include both the proper motion and the parallax during our period search,
ensuring that we recover the correct orbital period from the start. 

In this context, we note that the Lomb-Scargle periodogram is a special case of
Least Squares minimization (Cumming 2004), where the Least Squares minima (or 
the peaks of the periodogram) identify the correct periodicities when the true
signal is close to a sinusoid (see Frescura et al. 2008). This point was 
first made by Scargle (1982). A Lomb-Scargle periodogram works very well with radial velocity data, where the only relevant parameter not related to the periodic motion 
is a constant offset. This performance breaks down, however, for astrometric
data, due to the time dependence of the proper motion and the parallax. If the
true signal's period is well-sampled, the Lomb-Scargle periodogram provides an 
answer close to the correct one (Traub et al. 2009), but its statistical 
interpretation in terms of significance and confidence level is unclear. 
We are continuing to investigate the comparative performance of both approaches 
when applied to astrometric data, but for the purposes of this initial paper,
we limit ourselves to the Least Squares approach.

The purpose of the Least Squares periodogram is to find an initial set of
astrometric parameters that can be used as a first approximation for a
fully Keplerian solution that minimizes some merit function (i.e., $\chi^2$).
The Least Squares periodogram approach allows the weighting of each 
observation properly at the initial period-search level, and provides 
all the parameters of the best-fit circular orbit (via the Thiele-Innes elements), 
which can be used as initial values to solve the fully nonlinear Kepler
problem (e.g., Wright et al. 2009).

 If one calculates the
$\bar{\chi}^2$ as a function of the period (i.e., draw a periodogram, see
Figure 7), the minimum $\bar{\chi}^2$ is the best circular
model fitting the data. Currently, the number of measured parameters is
comparable to the number of unknowns, and so a large number of
artificial least squares minima with $\bar{\chi}^2$ much smaller than 1
appear in the periodogram. However, this periodogram does give
information about the most important orbital phases that have to be 
sampled in order to eliminate spurious signals, as shown in Figure 7.
The best period for NLTT 48256 is at $15.19$ days and has a semi-amplitude of
$1.2$ mas. However, the model clearly fits the data \textit{too well}, 
as shown in the phased representation of the best orbit in Figure 8 (top). 
Other minima in the periodogram are due to the
aliasing of the noise with the sampling cadence
(Figure 8, center) and the coupling of the astrometric
motion with natural periodicities, such as the parallax (Figure 8, bottom). 
External constraints can be used
to suppress such unrealistic least squares minima, however the best
strategy is simply taking more data and optimizing the cadence to 
minimize the effects of the discrete sampling cadence.

In general, the minimum number of observations needed to constrain 
with confidence an orbital model is a complicated function of the number of
observations, signal-to-noise ratio, spacing of the observational epochs,
orbital period and eccentricity, and the choice of the statistical tests used
to choose the best orbital model (e.g., Sozzetti 2005; Casertano et al. 
2008; Ford 2008; Cumming et al. 2008; Wright \& Howard 2009). 
However, a lower bound on the number
of observational epochs required to solve for a planetary companion can 
be determined simply from linear algebra theory: at least $n$ independent
observations are needed to solve a system with $n$ unknown factors. 
For a circular Keplerian orbit, there are a total of 10 free parameters in
the astrometric solution for the orbit, proper motion, and parallax, while for an
eccentric orbit, there are 2 more free parameters (eccentricity and the
argument of periastron), for a total of 12 free parameters. For a circular
orbit, then, at least 5 observational epochs in two coordinates (R.A. and
Dec.) are required for a solution, and at least 6 epochs for an eccentric
orbit. Clearly, such minimal solutions must be considered dubious, and
require additional epochs for their validity to be properly assessed. We
expect that at least 10 or 12 epochs (20 or 24 measurements in R.A. or
Dec.) will be necessary to constrain circular or eccentric orbits, respectively.
In cases where additional information is available (e.g., radial velocities, 
or catalog proper motions), these estimates might be relaxed somewhat. 

\section{STEPS Data Analysis Pipeline}

 The STEPS data reduction process (Pravdo et al. 2004) begins by 
extracting square regions containing the target and reference stars 
from the raw frames and organizing them into a single file, from which
positions for all stars are determined. The cross-correlation of the 
reference star positions relative to the target star is determined by the 
weighted slope of the phase of the Fourier Transform of the images 
after summing them into horizontal and vertical distributions. This 
algorithm is insensitive to the background level, robust against changes 
in the shape of the point spread function (PSF), 
and maintains a signal-to-noise ratio comparable 
to matched filtering. Centroiding is then performed, and a preliminary 
astrometric solution is obtained by fitting a conformal six-term 
(three per axis) transformation for each CCD frame to a reference 
frame. The transformation is then applied to the target star as well,
allowing the target star position to be measured relative to the surrounding 
reference stars. An automated program then searches for frames whose 
astrometric noise is above a user-defined threshold. Generally, the threshold 
can be set to be very high, because the major cause of unusable data 
is missing reference stars or selection of the wrong star. After 
removal of the bad frames, the conformal transformation is run again
to form an intermediate astrometric solution. 

 The DCR effect is proportional to the tangent of the 
zenith angle and leads to a linear shift in right ascension 
and a parabolic shift (relative to the meridian position)
in declination. The relative DCR coefficient for each star is determined
empirically by fitting the right ascension shift. A single coefficient 
for each star is defined as the weighted average of the nightly 
coefficients, used to adjust the centroid positions, and the conformal 
transformation is rerun once again. The position of the target star 
relative to the reference frame is then known for that night.

 The final STEPS processing step is to fit the motion of the target 
star to a model of the its parallax, proper motion, and radial velocity.
STEPS uses relative, rather than absolute, proper motions and parallaxes.
The USNO subroutine ASSTAR is used to compute the astrometric wobble,
and the NAIF subroutine CONICS is used to determine the position and 
velocity of a suspected companion from an assumed set of elliptic
orbital elements, which are then varied in order to find the best fit.

 The analysis of NLTT 48256 using the STEPS pipeline is in 
good agreement with the analysis obtained using the CAPSCam data 
reduction scheme. However, the intra-night scatter and corresponding 
single epoch uncertainties are larger using the STEPS pipeline. We 
attribute this to the STEPS centroiding approach, which requires a 
finer sampling of the stellar PSF than is obtained using CAPSCam-S -- the 
seeing-disk to pixel ratio is about 15 to 20 with the STEPS camera on the 
Palomar 5-m telescope, while the ratio is about 5 for CAPScam-S on 
the 2.5-m du Pont.

 It is also significant
that the reference stars used in the two analyses are not exactly the 
same. We attribute the discrepancy in the obtained proper motions to 
these differences. In spite of these differences, both astrometric 
solutions agree to within the errors (see Table \ref{table_solution}). A refined centroiding 
algorithm using an analytic approximation of the PSF is being 
implemented on the STEPS pipeline in order to achieve improved compatibility
with the CAPSCam pipeline. Table \ref{table_shifts} presents the R.A. and Dec offsets as a
function of time for those who may wish to try their own fit to this CAPSCam-S
data.

\section{Comparison With Other Ground-Based Programs}

 Considerable progress has been made in the adaptation of CCDs 
to the field of astrometry. Most notable is its application in an
astrometric instrument called STEPS, which has been used at Keck and at
the Palomar 5-m telesope (Pravdo \& Shaklan 1996). STEPS is able to achieve 
high precision (approximately 0.25 to 0.5 milliarcsec) on 30 minute 
exposures of intrinsically faint M dwarfs in fields with bright,
well-distributed reference stars. STEPS has detected a low-mass M6-M8 
binary companion to the M dwarf GJ 164, with a noise floor of 
$\sim$ 1 milliarcsec (Pravdo et al. 2004). STEPS has also found
low mass companions to the M dwarfs GJ 802 (Pravdo, Shaklan,
\& Lloyd 2005) and G78-28AB and GJ 231.1BC (Pravdo et al. 2006).
Most recently, STEPS appears to have achieved the first astrometric
discovery of a planet, finding evidence for a $\sim$ 6 Jupiter mass 
planet candidate with an orbital period of 0.744 year around the 
M8 star VB10 (Pravdo \& Shaklan 2009).

 Bartlett, Ianna, \& Begam (2009) summarized the results of their three 
years of observations of thirteen nearby (within 25 pc) stars in the University
of Virginia Southern Parallax Program. They were able to achieve parallaxes
with errors of less than 3 milliarcsec for these stars, all early 
M dwarfs, using a minimum of at least 50 observational epochs. One 
star, LHS 288, appears to exhibit a wobble indicative of a very low 
mass companion, and this star appears in the CAPSCam-S observing list as well.

 Henry et al. (1997, 2006) have been prolific in finding the closest stars
and in searching the 91 closest red and white dwarf stars for low mass
companions, using 1m-class telescopes and existing CCDs at Cerro 
Tololo (ASPENS). The ASPENS program typically achieves astrometric 
accuracies of about 1.5 milliarcsecs. The median seeing at Cerro Tololo 
is about 0.95 arcsec, compared to 0.6 arcsec on Las 
Campanas (Persson et al. 1990), both measured at 0.5 microns. As a result,
and because of the larger aperture (2.5-m), better plate scale, and
smaller optical distortions on the du Pont, we expect our survey
to achieve better long-term astrometric accuracies.

 Bower et al. (2009) used the Very Large Array (VLA) to perform a radio
astrometry survey of 172 active M dwarfs within 10 pc. They were able
to place upper bounds on the presence of any planetary companions of
no more than 3 to 6 Jupiter masses orbiting at $\sim$ 1 AU. The VLA survey
typically yielded milliarcsecond resolution, though for four stars that
were observed at multiple epochs, the RMS residuals were $\sim$ 0.2 
milliarcsecs.

 Adaptive optics (AO) has been used on the Palomar 5-m telescope to improve
the astrometric performance, with accuracies achieved as good as 
0.1 milliarcsec over 2 months (Cameron, Britton, \& Kulkarni 2009).
The FORS1 camera on one of the Very Large Telescope (VLT) 8-m telescopes
has achieved astrometric accuracies of 0.2 to 0.3 milliarcsecs
per single measurement (Lazorenko et al. 2007).

 Interferometry has the greatest ultimate potential for astrometric 
detection of extrasolar planets (Shao \& Colavita 1992). The Palomar
Testbed Interferometer (PTI), with 0.4-m aperture siderostats, 
has been used primarily for interferometer 
instrumentation and software development (Colavita \& Shao 1994). 
The main limitation of the PTI is that it can only use one reference
star at a time, and that one star must be quite bright and very near to
the target. Thus the PTI does not have a significant planet
survey capacity, even though it has demonstrated an astrometric
accuracy of about 0.1 milliarcseconds over a period of a few days.
The PTI has been successfully employed in very high precision studies 
of several double stars  (e.g., Boden et al. 1999; Boden \& Lane 2001).

 The Palomar AO, VLT/FORS1, and PTI experiments have thus demonstrated high  
precision, but only over very short time baselines (days to months) compared to  
our results to date (spanning several years of CAPSCam-S data). Given
the need for long term stability (by definition) in an astrometric
planet search, we believe that CAPSCam is well-suited for making
significant astrometric discoveries.

 The Very Large Telescope Interferometer (VLTI) will also be capable 
of astrometric planet searches. The VLTI plans to begin astrometric observations 
in 2010 with four auxiliary (1.8-m) telescopes forming the PRIMA interferometer 
and to achieve an astrometric accuracy of about 0.03 milliarcsecs. 
However, late M, L, and T dwarfs will be too faint for interferometric 
astrometry on $\sim$ 100-m baselines with 1.8-m telescopes and with
bandpasses limited by the difficulty of achieving achromatic fringes. 
In fact, the PRIMA planet search of 100 to 200 stars is planned only
for target stars that are brighter than K = 12 to 14 (J. Setiawan,
2004, private communication), limiting this search to stars earlier
than early M type. 

 Radial velocity surveys have been by far the most successful and
valuable of all the planet-search techniques. However, these spectroscopic
searches typically require a high flux of optical stellar photons in order 
to achieve velocity precisions of a few m/sec, and consequently they have been 
generally limited to target stars brighter than about V = 11 
(R. P. Butler, private communication, 2008). Our planned 
astrometric search will be complementary to the spectroscopic surveys 
of low mass stars (e.g., Butler et al. 2004), as our
targets are typically no brighter than V = 13. For example, the HARPS  
GTO program (PI: Michel Mayor) on the La Silla 3.6-m telescope includes 
only 10 stars with spectral types later than M5 and none as late as M7.
The Carnegie-California search has a similar number of very late 
type stars (Wright et al. 2004). While radial
velocity surveys are intrinsically better suited for discovering
planets in short-period orbits (Figure 9), astrometry is
better suited for long-period orbits, requiring that the CAPSCam
planet search be at least a decade-long effort. Ideally, 
astrometry and radial velocity could observe the same target
stars, so that astrometry could remove orbital inclination 
ambiguity of radial velocity observations. A few planets have been 
discovered by radial velocity that are close enough, faint enough, 
and located far enough south to be prime candidates for CAPSCam astrometry.

\section{Comparison With Space-Based Programs}

 The European Space Agency (ESA) Hipparcos satellite (1989-93) and the
resultant catalog comprise the most successful astrometric
effort in history. Hipparcos obtained the parallaxes of nearly 120,000 stars 
with a median precision of about 1 milliarcsecond. Hipparcos' ability to 
detect an astrometric planet perturbation was limited primarily by the 
mission's duration of only 3.36 years and by its annual precision 
of approximately 2 milliarcseconds.

 Astrometric efforts with the Hubble Space Telescope (HST) have been almost 
entirely confined to use of the interferometers of the fine guidance system
(FGS) instead of HST's CCD cameras.  The resulting studies are among 
the highest precision (approximately 0.5 milliarcsec) planet searches to date
(Benedict et al. 1999). Unfortunately these searches were of
limited duration. The FGS has also been used to place an upper limit
on the mass of the short-period companion to 55 Rho$^1$ Cancri of
about 30 $M_{Jup}$ (McGrath et al. 2002), a result that differed
considerably from the Hipparcos evidence for a wobble large enough
to require the presence of an M dwarf companion with a mass of 126 $M_{Jup}$. 
The FGS measurements, with an accuracy of about 0.3 milliarcseconds, 
rule out the Hipparcos claim. More importantly, Benedict et al. (2002)
used the FGS to determine the mass of the outermost planet of the 
GJ 876 system to be $1.9 M_{Jup}$, the first time that astrometry
has determined the mass of an extrasolar planet. The low mass of
the GJ 876 primary star ($0.32 M_\odot$) enabled this detection,
along with knowing basic orbital parameters from the original
spectroscopic detection of the star's planets. While the FGS evidently
can be a potent astrometric instrument, the difficulty of obtaining 
precious HST time for lengthy astrometric surveys limits
its use to following up on particularly promising spectroscopic 
detections, such as GJ 876.

\section{Conclusions}

 Our analysis of the target star NLTT 48256 shows that an 
astrometric accuracy better than $0.4$ mas can be obtained over
a time scale of several years with CAPSCam-S on the du Pont 2.5-m.
Our preliminary DCR correction indicates that full differential 
chromatic corrections will be required in our data pipeline if 
accuracies below $0.3$ mas/epoch are to be achieved. We are presently
working in this direction. Such a long-term accuracy would
probably be the best accuracy that could be achieved with seeing-limited
observations over a time-scale of many years on a 2.5-m-telescope,
and would be the result of the combination of a very stable instrument 
with a strict data processing scheme.

 A Jupiter-mass planet on a Jupiter-like orbit around a solar-mass
star produces an astrometric wobble of the primary star of 2 milliarcsec 
(measured peak to valley) when viewed from a distance of 5 pc. Thus an 
astrometric accuracy of (plus or minus) 0.25 milliarcsec would allow the
detection of such a planetary companion with a signal-to-noise
ratio of 4. For lower mass stars, the nominal astrometric detection limit 
drops to proportionately lower planet masses. Brown dwarf companions 
at similar orbital distances will be considerably easier to discover.

 Optical Doppler surveys have observed about 200 different M-type stars. 
These are mainly of spectral type earlier than M3, due to the constraint of  
needing high V-band magnitudes. Cummings et al. (2008) find a  
statistically significant deficit of planets with masses between 0.3
and 10 Jupiter masses and with orbital periods $<$ 2000 d around M-dwarfs 
relative to FGK-dwarfs -- a gas giant planet fraction of 2\% rather  
than 7.5\% for these relatively short ($\sim$ 5 yr) orbital periods.
Investigations of the period-mass relation for M stars are  
still impossible due to the small number of detected planets.  The  
planet fraction around stars and brown dwarfs with mass $< 0.25 M_ 
\odot$ is largely unprobed. Doppler searches of small samples of 
young stars have so far revealed an 18 Jupiter-mass object around a 
brown dwarf (Cha H$\alpha$ 8; Joergens \& Muller 2008). Optical  
Doppler surveys have targeted very few of these faint stars because  
integration times are prohibitively large, and infrared Doppler surveys are in  
their infancy. Direct imaging studies can only search for planets widely  
separated from young brown dwarfs (e.g., 2MASS1207; Chauvin et al. 2004). 
Astrometry searches closer in, and removes the orbital inclination ambiguity 
of Doppler surveys. Hence we believe that by targeting late M, L, and T 
dwarfs for astrometric monitoring for a decade or more, CAPSCam will
make an important contribution to the census of planetary systems.

 We believe that a sample of around 100 stars is sufficiently 
large to ensure a reasonable statistical measure of the frequency of 
long-period gas giant planets (and binary companions) around late M 
and later type dwarfs, companions with orbital periods long enough 
to permit habitable rocky planets to orbit these stars on shorter 
period orbits. M dwarfs have recently been recognized as attractive 
targets in the search for life beyond the Solar System 
(Segura et al. 2005; Tarter et al. 2007), yet late M dwarfs
are not being studied by optical Doppler surveys in any great number.
While CAPSCam will not 
be able to detect habitable terrestrial planets, we will be able
to point the way for searches by future ground- and space-based
telescopes designed to discover new Earths around the closest
stars. In fact, the report of the Exoplanet Task Force 
(Lunine et al. 2008) specifically calls for planet searches
around M dwarfs to be a fast-track effort for ground-based
and existing space telescopes (see their Figure 1).

 We thank Paul Butler, Sandy Keiser, and Dave Monet for their key 
contributions to this effort, and Wendy Freedman, Mark Phillips, and Miguel 
Roth for their steady support of this ambitious program at Las Campanas. 
Oscar Duhalde, Javier Fuentes, Gast\'on Guti\'errez, Herman Olivares, 
David Osip, Fernando Peralta, Frank P\'erez, Patricio Pinto, and 
Andr\'es Rivera have provided valuable assistance at Las Campanas. We 
thank the referee as well, whose comments have helped to improve the 
paper. This work has been supported in part by NSF grants AST-0352912 and 
AST-0305913, NASA Planetary Geology and Geophysics grant NNX07AP46G, 
NASA Origins of Solar Systems grant NNG05GI10G, and NASA Astrobiology 
Institute grant NCC2-1056. This research has made use of the SIMBAD 
database, operated at CDS, Strasbourg, France.

\vfill\eject

\begin{deluxetable}{lllll}
\tablecaption{CAPSCam array readout times (in seconds).\label{table_readout}}
\tablewidth{0pt}
\tablehead{\multicolumn{5}{c}{FF times (2048$\times$2048)}}
\startdata
FF-flush time:  &2.146 &           &          & \\
FF-read time:  &8.397  &           &          & \\
\cutinhead{\ \ \ \ \ \ \ \ \ \ \ \ \ \ \ \ \ \ \ \ \  \ \ \ \ \ \ \ \ \ \ GW times}
GW-pixel size:    &32x32   &64x64    &128x128   &256x256\\
\hline
GW-flush time: &0.003   &0.010    & 0.036    & 0.137\\
GW-read time:  &0.009   &0.034    & 0.133    & 0.528\\
\enddata
\end{deluxetable}

\suppressfloats
\begin{deluxetable}{ll}
\tablecaption{CAPSCam sample exposure sequence for 60 s FF, 1 s GW (64$\times$64).\label{table_exposure}}
\tablewidth{0pt}
\tablehead{\colhead{time (s)} &\colhead{Action}}
\tabletypesize{\scriptsize}
\startdata
 0.000   & FF flush start \\
 2.146   & FF 1st read start \\
10.543   & FF read done (wait until next full second) \\
         & \\
11.000   & GW flush start \\
11.010   & GW 1st read start (exposure 1) \\
11.044   & open shutter \\
12.044   & close shutter, wait 50 ms \\
12.094   & GW 2nd read start (exposure 1) \\
12.128   & GW read done (wait until next 0.1 second) \\
         &\\
12.200   & GW flush start \\
12.210   & GW 1st read start (exposure 2) \\
12.244   & open shutter \\
13.244   & close shutter, wait 50 ms \\
13.294   & GW 2nd read start (exposure 1) \\
13.328   & GW read done (wait until next 0.1 second) \\
         & \\
13.400   & GW flush start \\
...      & elapsed time = $11+60*1.2 = 83$ \\
         &\\
83.000   & 50ms wait for shutter to close  \\
         &\\
83.050   &  FF 2nd read start \\
91.477   & FF 2nd read done  \\
\enddata
\end{deluxetable}

\begin{deluxetable}{lc}
\tablecaption{Catalog information on NLTT 48256. 
Proper motion and coordinates are
from the NLTT catalog (Salim \& Gould 
2003) and color is from the NOMAD catalog
(USNO-B1 + 2MASS, Zacharias et al. 2005).\label{table_nltt48256}}
\tablewidth{0pt}
\tablehead{
\multicolumn{2}{c}{Reference epoch JD 2000}}
\startdata
RA       &19   48 37.5\\
Dec      &$-$19 32 14.3\\
$\mu$RA  &$-$38  $\pm$ 20 mas/yr \\
$\mu$Dec &$-$187 $\pm$ 20 mas/yr\\
\hline
R        &16.84\\
R$-$J    &3.63 \\
\enddata
\end{deluxetable}

\begin{deluxetable}{lccccc}
\tablecaption{Log of observations of Field 453. The SNR on NLTT 48256 is
always over 1000 and the faintest reference frame star has a typical SNR of
200-300. Exposure times are given in seconds while the seeing is in 
arcseconds. \label{table_obs}}
\tablewidth{0pt}
\tablehead{
\colhead{Date} &\colhead{FF Exptime}
                     &\colhead{GW Exptime}
                           &\colhead{\# FF Images}
                                &\colhead{Seeing}
                                      &\colhead{Airmass}
                                             }
\startdata
2007 Jul 04    &60   &--   &90  &0.9  &1.02   \\
2007 Jul 08    &100  &--   &36  &1.2  &1.02  \\
2007 Aug 31    &60   &--   &60  &0.9  &1.01 \\
2008 Jul 14    &30   &--   &80  &0.8  &1.03  \\
2008 Sep 12    &30   &--   &80  &1.0  &1.16 \\
2009 Apr 10    &60   &--   &26  &0.7  &1.14 \\
               &45   &--   &15  &0.7  &1.06 \\
2009 Jun 01    &120   &30   &20  &0.9  &1.02 \\
2009 Jun 04    &120   &30   &20  &0.8  &1.02 \\

\enddata
\end{deluxetable}

\begin{deluxetable}{lcccc}
\tablecaption{Astrometric solution for NLTT 48256. Reference epoch is 
August 31, 2008 or Julian Date: 2454343.558125. In all cases there are 5
effective epochs and 366 images.\label{table_solution}}
\tablewidth{0pt}
\tablehead{
             &\multicolumn{2}{c}{CAPSCam\tablenotemark{a}} &\multicolumn{2}{c}{STEPS\tablenotemark{b}}\\
             &\colhead{Achromatic}
                                     &\colhead{Chromatic}
                                             &}
\startdata
RA            &19 48 37.488      &19 48   37.488	 &  &  \\
Dec           &$-$19 32 15.926   &$-$19 32 15.925	 &  &  \\      
muRA(mas/yr)  &$-$40.04 $\pm$ 0.13 &$-$39.99  $\pm$ 0.12  &$-$41.7  &(+1.10 $-$1.00)  \\
muDE(mas/yr)  &$-$187.71$\pm$ 0.13 &$-$187.66 $\pm$ 0.12  &$-$188.8 &(+1.10,$-$1.00) \\
par(mas)      &17.80    $\pm$ 0.15 &16.69     $\pm$ 0.14  &17.37    &(+0.49,$-$0.46)  \\
RMS(mas/epoch)&    0.38            &0.35                  &1.50     &\\
\enddata
\tablenotetext{a}{Parameter uncertainties in CAPSCam are given as the standard 
     deviations obtained from the covariance matrix of the linearized 
     least squares solution.}
\tablenotetext{b}{Parameter uncertainties in STEPS are given as the interval 
     with a 68\% confidence level.}
\end{deluxetable}

\begin{deluxetable}{lllcccc}
\tablecaption{Astrometric shifts of NLTT 48256 in local plane coordinates, 
uncertainties, and postfit residuals after fitting an offset, parallax, 
and proper motion only. All angles given in milliarcseconds (mas).\label{table_shifts}}
\tablewidth{0pt}
\tablehead{
\colhead{Julian Date} &\colhead{RA Shift}
                            &\colhead{Dec Shift}
                                         &\colhead{RA unc} 
                                                  &\colhead{Dec unc}
                                                           &\colhead{O-C RA} &\colhead{O-C Dec}}
\startdata
2454285.726960 & -1.566825 &  -3.984330  &0.143    &0.205   &-0.008    &-0.801 \\
2454289.733512 & -2.872852 &  -5.162385  &0.304    &0.383   & 0.187    & 0.271 \\
2454343.589780 &-22.608162 & -35.169847  &0.184    &0.263   &-0.117    & 0.566 \\
2454661.673801 &-45.457509 &-197.023902  &0.085    &0.259   &-0.030    &-0.054 \\
2454722.625970 &-66.343008 &-230.239554  &0.218    &0.472   &-0.070    & 0.883 \\
2454928.216687 &-59.445586 &-332.217484  &0.295    &0.335   & 0.090    &-0.472 \\
\enddata
\end{deluxetable}

\suppressfloats
   
\begin{figure}
\vspace{-0.5in}
\plotone{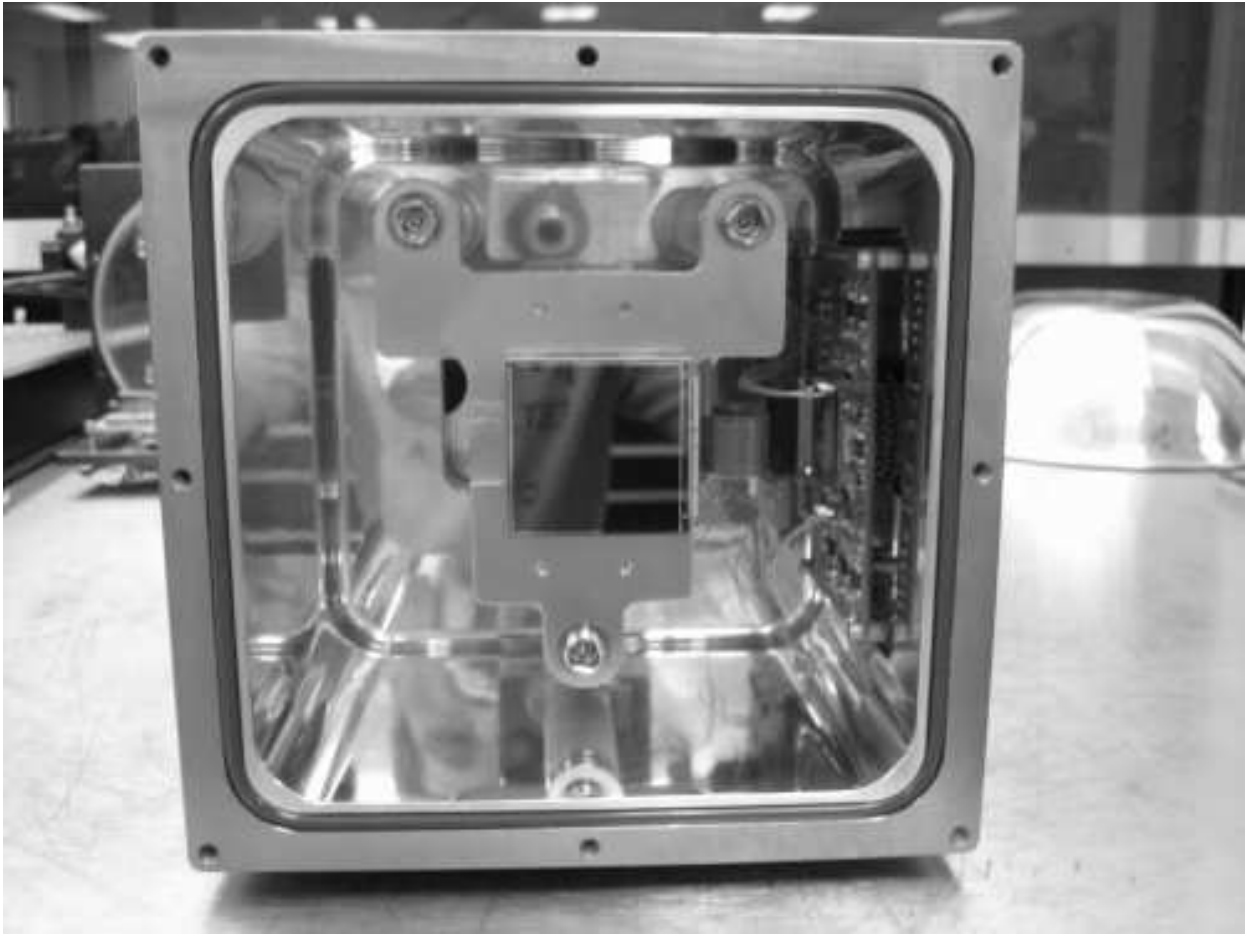}
\caption{The Teledyne Hawaii-2RG array can be seen mounted in the
center of CAPSCam-N, shown here with the Barr Associates filter/window 
removed. The Hawaii-2RG array is approximately 2" by 2" in size.}
\end{figure}

\suppressfloats

\begin{figure}
\vspace{-0.5in}
\plotone{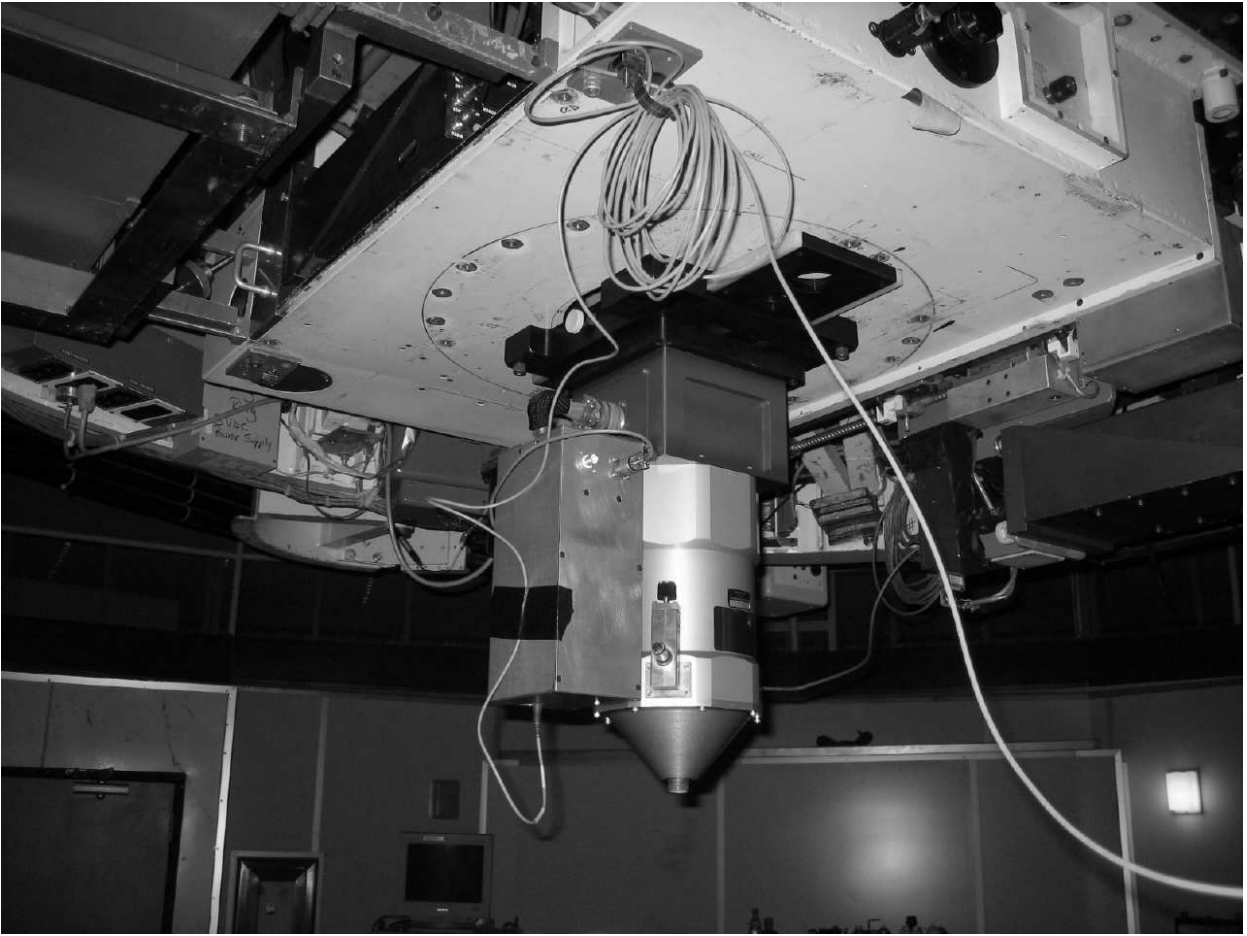}
\caption{CAPSCam-S is shown mounted at the Cassegrain focus of the
2.5-m du Pont telescope at Carnegie's Las Campanas Observatory
in Chile.}
\end{figure}

\begin{figure}
\vspace{-0.5in}
\plotone{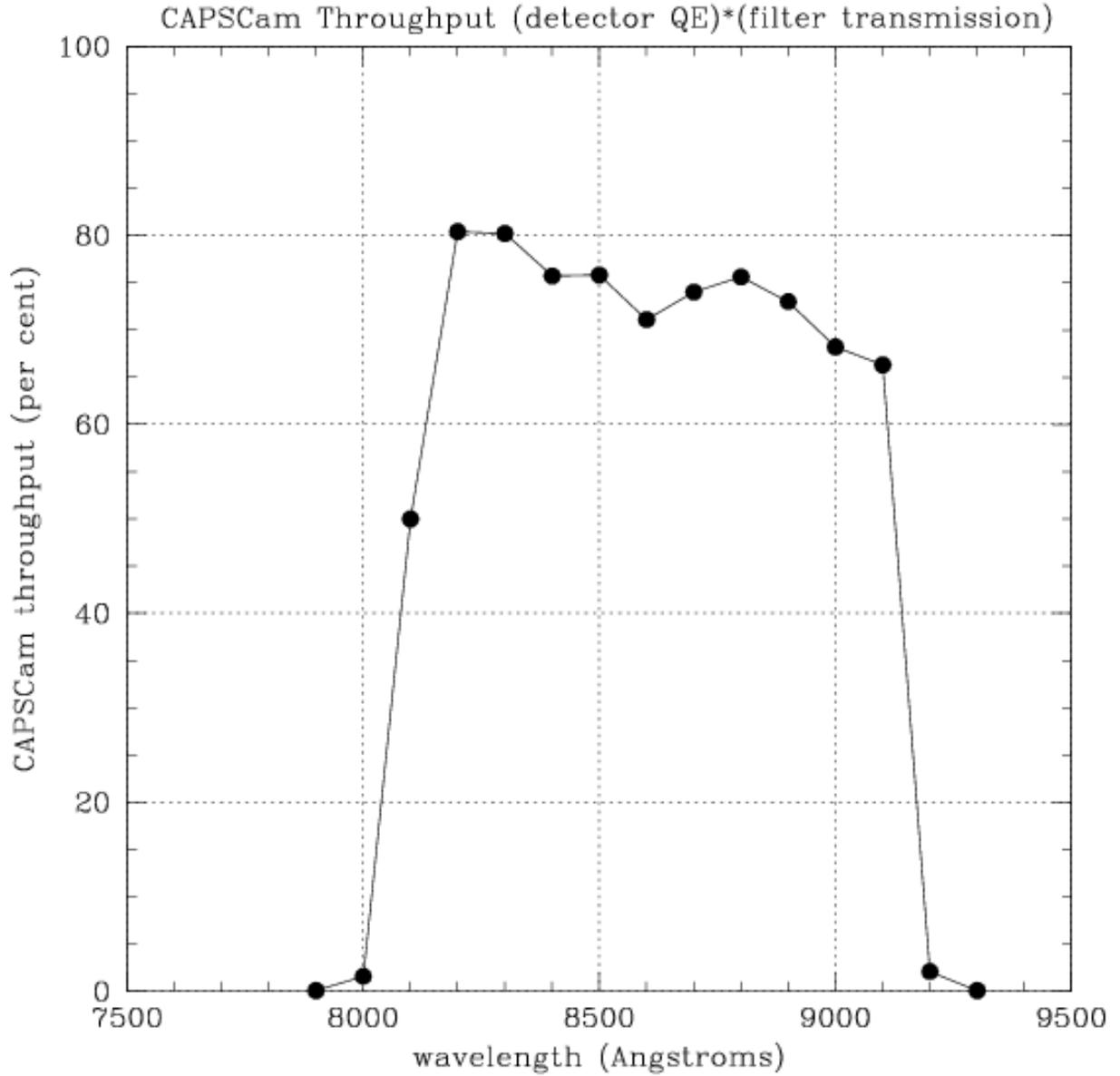}
\caption{CAPSCam-S throughput in percent as a function of wavelength. The 
throughput is the product of the quantum efficiency of the Hawaii-2RG
detector and the transmission function of the filter/window. 
CAPSCam is optimized for the study of M dwarf stars, with a 
bandpass of about 100 nanometers centered at about 865 nanometers.}
\end{figure}

\suppressfloats

\begin{figure}
\center
\includegraphics[angle=0, width=5in,clip]{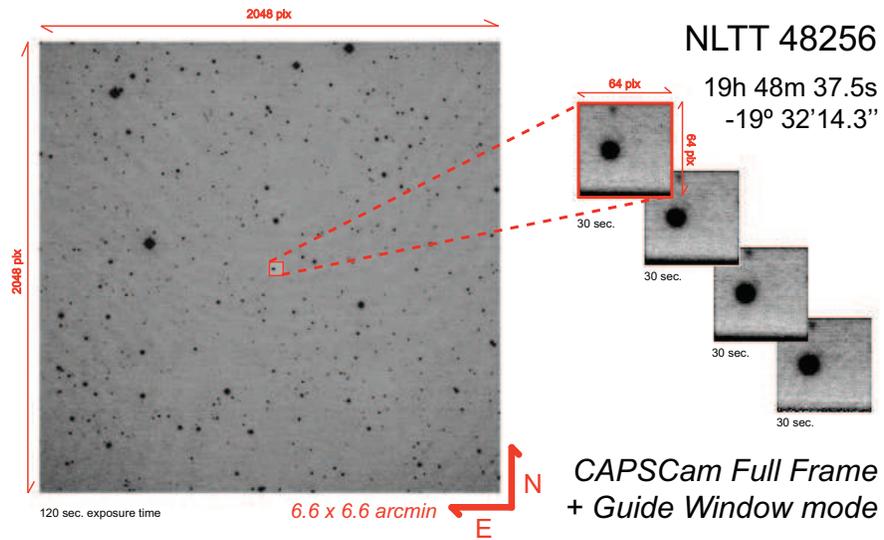}
\caption{CAPSCam-S Full Frame image of Field 453, with NLTT 48256 located 
in the Guide Window.}
\end{figure}

\begin{figure}
\center
\includegraphics[angle=0, width=5in,clip]{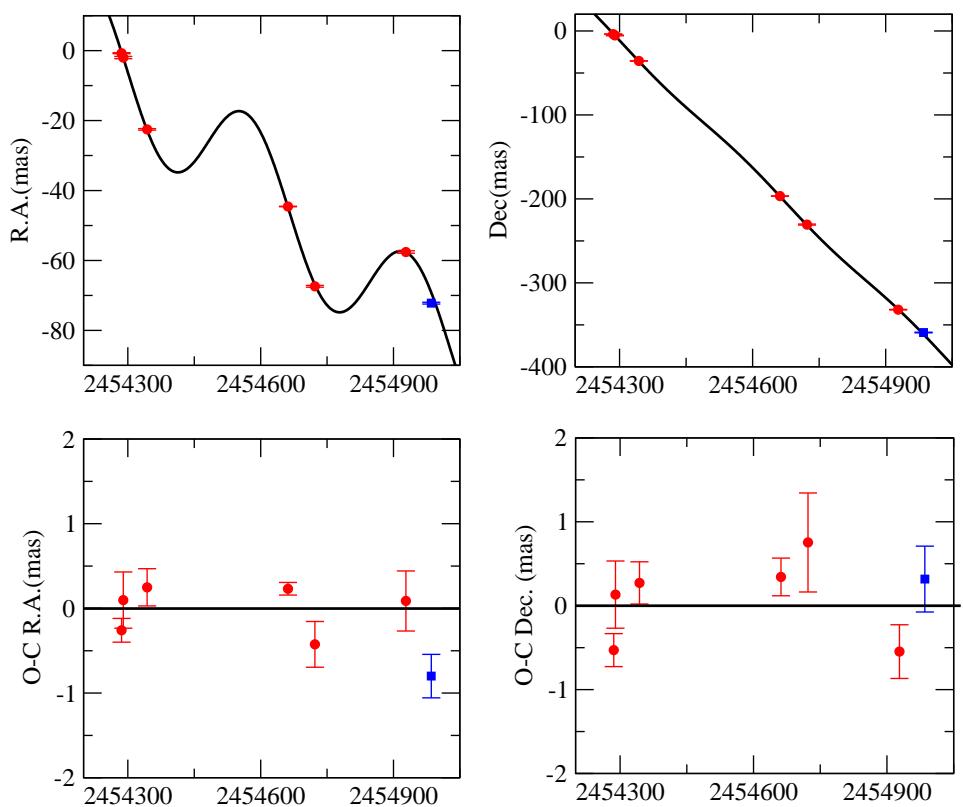}
\caption{ 
\textbf{Top.} R.A. (left) and Declination (right) shifts as a function of 
time for NLTT 48256. The parallax wobble can be clearly seen on the R.A. motion.
\textbf{Bottom.} R.A. and Declination residuals (observed minus computed) 
with respect to the best fit model. All vertical axes are in milliarseconds (mas).
The right-most data point is based on Guide Window and Full Frame data, while 
the others are Full Frame only. Use of the GW does not introduce any significant
bias or jitter to the data.}
\end{figure}

\suppressfloats

\begin{figure}
\center
\includegraphics[angle=0, width=5in, clip]{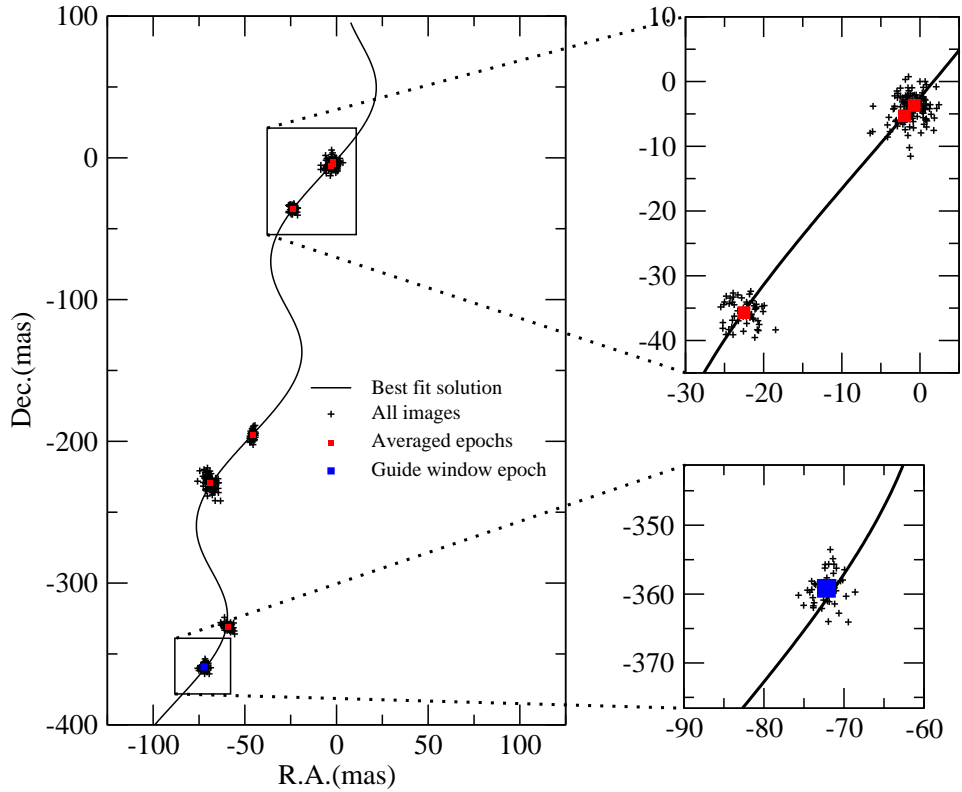}
\caption{Motion of NLTT 48256 on the sky (R.A. vs. Dec.) in mas. On 
the top right, a zoom-in of the first two epochs is shown. The small crosses are the 
positions as measured from each individual image. The scatter is 
consistent with a standard deviation of $\sim 2$ mas per image in each 
direction, which is $\sim$ 1/100th of the pixel size of 0.194 arcsec. The
bottom-most data point (and zoom-in) is based on Guide Window and Full Frame 
data, while the others are Full Frame only.}
\end{figure}

\suppressfloats

\begin{figure}
\center
\includegraphics[angle=0, width=6in, clip]{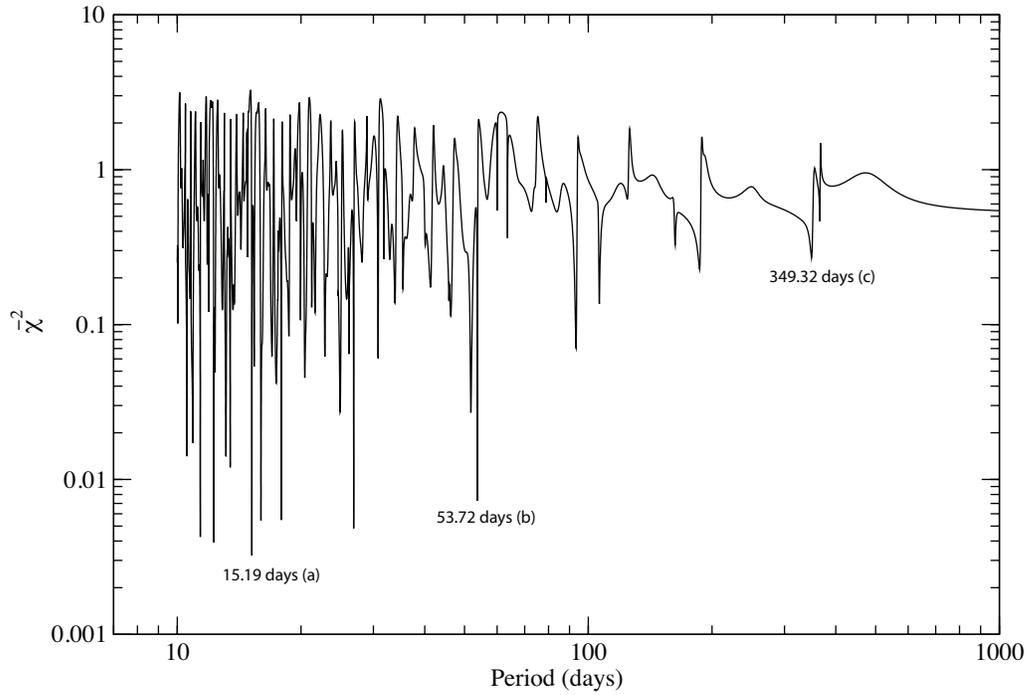}
\vspace{0.5in}
\caption{Best $\bar{\chi}^2$ as a function of the period (periodogram)
for NLTT 48256. The
minima represent the candidate periods. Since the number of observations is
still small, most of the minima are spurious and can be attributed to 
well-known issues related to poor data sampling (see Figure 8).}
\end{figure}

\begin{figure}
\center
\includegraphics[angle=0, width=5in, clip]{f8a.eps}
\includegraphics[angle=0, width=5in, clip]{f8b.eps}
\includegraphics[angle=0, width=5in, clip]{f8c.eps}
\caption{False minima in the periodogram for NLTT 48256 can lead to spurious solutions. 
\textbf{Top.} Best fit period, showing R.A. and Dec. as functions of the 
orbital phase, for $P \sim 15$ days. The $\bar{\chi}^2$ is much smaller
than 1 because the number of parameters being fitted almost matches the 
number of observations.
\textbf{Center.} Aliasing with the observing cadence ($P \sim 54$ days). 
All the data appears to be concentrated on a small orbital phase.
\textbf{Bottom.} Coupling with the parallax. Here the apparent period is 
$P \sim 349$ days, about one Earth year. The parallax obtained for this 
solution is $-25$ mas, clearly illustrating that this spurious signal 
is due to the strong correlation with the parallax period.}
\end{figure}

\suppressfloats
   
\begin{figure}
\vspace{-1.0in}
\plotone{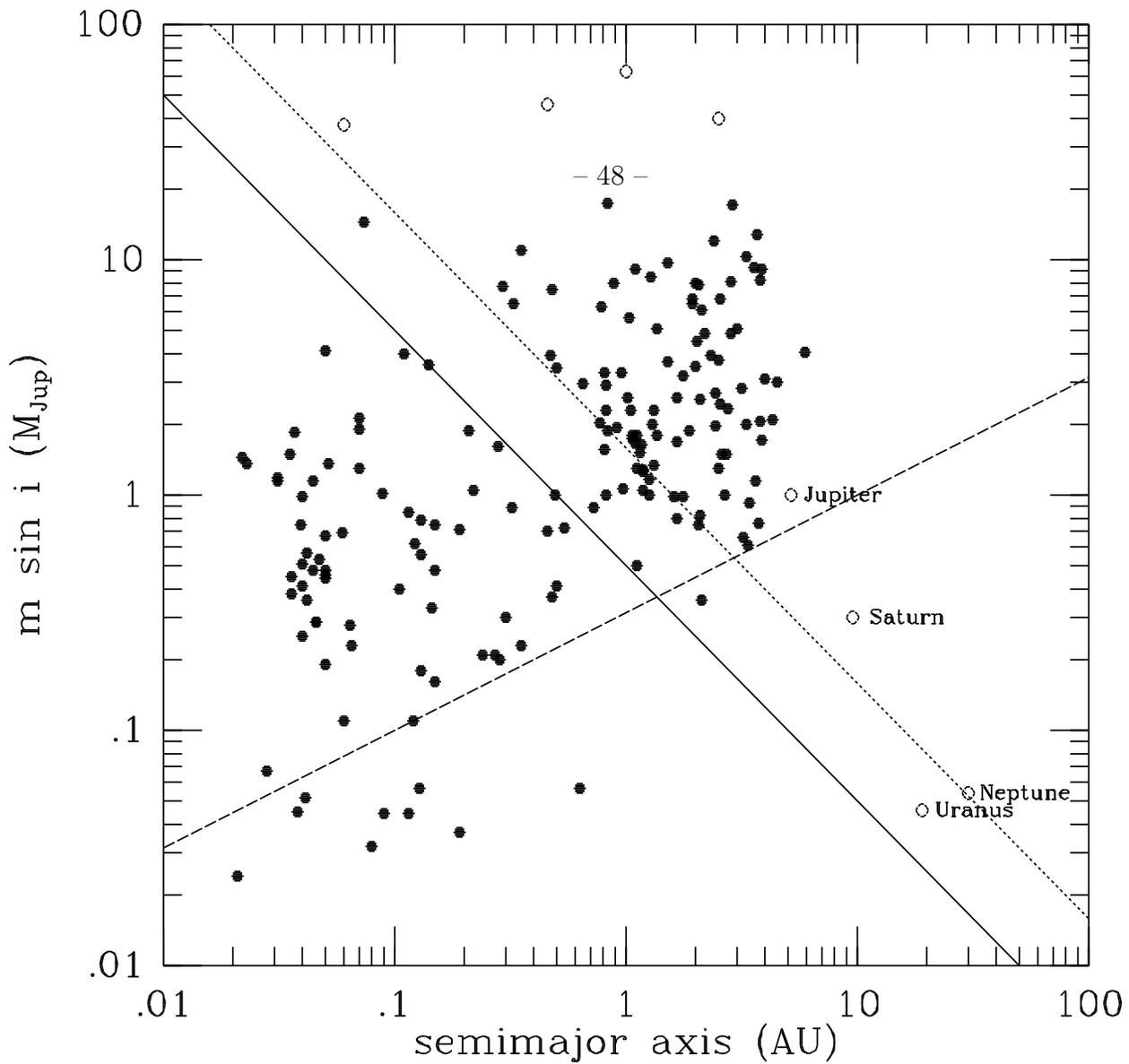}
\caption{Discovery space for gas giants and hot super-Earths (solid dots)
and brown dwarfs (unlabelled open circles) found by Doppler spectroscopy.
Data points plotted are from the IAU Working Group on Extrasolar Planets
(WGESP) web site (http://www.dtm.ciw.edu/users/boss/planets.html), which is
complete up to August 2006, augmented with the most recent hot super-Earths.
Nearly all of the host stars are G dwarfs, though the hot super-Earths
tend to be found around early M dwarfs. Objects with masses above
13 Jupiter masses are considered by the WGESP to be brown dwarfs, not planets,
though this plot represents objects with m sin $i$ values below 20 Jupiter
masses as planets. The oblique dashed line illustrates the dependence of 
the sensitivity limit for spectroscopic detections on semimajor axis (for a 
signal-to-noise ratio of four for 2.5 m/s Doppler precision and a solar-mass
host star), while the oblique 
solid and dotted lines represent the limits for astrometric detections with 
0.25 milliarcsec accuracy and a signal to noise ratio of four, for a 
late M dwarf target star at distances of 5 pc and 15 pc, respectively.}
\end{figure}

\end{document}